\documentclass[english,aps,prl,amsmath,amssymb,reprint,superscriptaddress]{revtex4-1}
\usepackage[T1]{fontenc}
\usepackage{textcomp}
\usepackage[utf8]{inputenc}
\usepackage{color}
\usepackage{babel}
\usepackage{amstext}
\usepackage{amssymb}
\usepackage{graphicx}
\usepackage[bookmarks=true,bookmarksnumbered=false,bookmarksopen=false,
 breaklinks=false,pdfborder={0 0 0},pdfborderstyle={},backref=false,colorlinks=true,pdfpagemode=FullScreen]
 {hyperref}
\hypersetup{pdftitle={Dynamic buckling of Si tetramers on the Si(111)-7x7 surface},
 pdfauthor={R. Zhachuk, J. Coutinho and D. Sheglov},
 pdfkeywords={Silicon, Surface, Structure, STM, DFT},
 allcolors=blue}


\begin{document}
\title{Dynamic buckling of Si tetramers on the Si$(111)\textrm{-}7\times7$
surface}
\author{R. Zhachuk}
\email{zhachuk@gmail.com}

\affiliation{Department of Physics \& i3N, University of Aveiro, Campus Santiago,
3810-193 Aveiro, Portugal}
\author{J. Coutinho}
\affiliation{Department of Physics \& i3N, University of Aveiro, Campus Santiago,
3810-193 Aveiro, Portugal}
\author{D. Sheglov}
\affiliation{Institute of Semiconductor Physics, pr. Lavrentyeva 13, Novosibirsk
630090, Russia}
\date{\today}
\begin{abstract}
The atomic structure of Si tetramers that form on the Si$(111)\textrm{-}7\times7$
surface during homoepitaxy, is investigated by means of first principles
calculations with the currently available atomistic model as starting
point. It is demonstrated that the rectangular shape of the Si tetramer
is unstable against buckling. Comparison of calculated results with
available scanning tunneling microscopy (STM) data provides a new
understanding of the problem, indicating that the recorded STM images
are influenced by dynamic buckling.

\noindent \emph{This article may be downloaded for personal use only. Any other use requires prior permission of the author and AIP Publishing. This article appeared in J. Chem. Phys. 162, 234704 (2025) and may be found at \href{https://doi.org/10.1063/5.0273497}{https://doi.org/10.1063/5.0273497}. {\copyright} 2025 Author(s). This article is distributed under a Creative Commons Attribution-Noncommercial-ShareAlike (CC BY-NC-SA) License.}
\end{abstract}


\maketitle

\section{Introduction}

\emph{Magic clusters} can form both in free space \citep{kni84,bra97}
as well as on sample surfaces \citep{man98,wan08}. They exhibit extraordinary
stability and contain a specific (\emph{magic}) number of atoms arranged
in a well-defined structure, often differing both from the bulk material
and surface reconstruction geometries. The reason for the enhanced
stability of magic clusters on silicon surface is a local minimum
of energy for certain groups of atoms, which display close-to-optimal
bond topology of their structure \citep{wan08}. Much interest on
magic clusters formed on Si surfaces stems from their uniform size
and shape, making them promising candidates for surface nanopatterning
of nanoelectronic and photonic devices. Additional interest comes
from their influence on the Si homoepitaxy. They act as traps for
diffusing adatoms, and it has been reported that some Si magic clusters
on Si$(111)$ surface may even migrate as a whole, retaining their
atomic geometry along the way \citep{hwa99}. Such a picture could
significantly change our understanding of crystal growth which is
mainly based on the classic Burton-Cabrera-Frank theory \citep{bcf51,uwa16},
with repercussions on fundamental mechanisms such as step fluctuations
or island formation \citep{hwa99,hwa02,ho04}.

In Si homoepitaxy, two types of magic clusters stand out from other
structures: tetramers (referred to as C-structures by Tanaka \emph{et
al.} \citep{tan94}) and hexamers (observed by Hwang \emph{et al.}
by STM \citep{hwa99}). The atomic model of Si tetramers, which form
at the boundary between two Si$(111)\textrm{-}7\times7$ structure
halves is shown in Figs.~\ref{fig1}(a) and (b). In its original
form, the model offered a qualitative explanation for the scanning
tunneling microscopy (STM) images originating from the clusters shown
in Figs.~\ref{fig1}(c) and (d) \citep{tan94}. However, (i) their
stability was never explored, and (ii) the proposed explanation for
the observed STM images is questionable in light of previous work
that demonstrated the instability of symmetric dimers observed on
Si$(100)$ surface \citep{cha79}.

In this paper, we employ first principles density functional theory
(DFT) calculations to investigate the atomic structure of Si tetramers
on Si$(111)\textrm{-}7\times7$ surface \citep{tan94,sat00,sat00b,ho04}.
We demonstrate that the geometry of the Si tetramers, as proposed
in Ref.~\onlinecite{tan94}, is unstable against buckling. Instead,
we shown that the experimental STM images of tetramers result from
dynamic flip-flop motion of a four-member ring of Si atoms, alternating
between two stable buckled states.

\section{Calculation details}

Ground state geometries and electronic structures were found using
the $\mathtt{SIESTA}$ software \citep{sol02}. Here the Kohn-Sham
wave functions were described with help of linear combinations of
atom-centered orbitals of the Sankey-Niklewski type, which included
multiple zeta orbitals and polarization functions (13 functions per
Si atom). The electron density and potential terms were calculated
on a real-space grid with spacing equivalent to a plane-wave cutoff
of $200\,\mathrm{Ry}$. For the evaluation of energy barriers between
stable structures we employed the nudged elastic band (NEB) method
as implemented in the planewave-based $\mathtt{VASP}$ code \citep{kre96,kre96a}.
In this case a 250~eV planewave cut-off was imposed for the description
of the valence states. All calculations were performed using a recipe
of the generalized gradient approximation to the exchange and correlation
interactions between electrons, which is commonly referred to as PBEsol
\citep{per08}.

\begin{figure}
\includegraphics[clip,width=8cm]{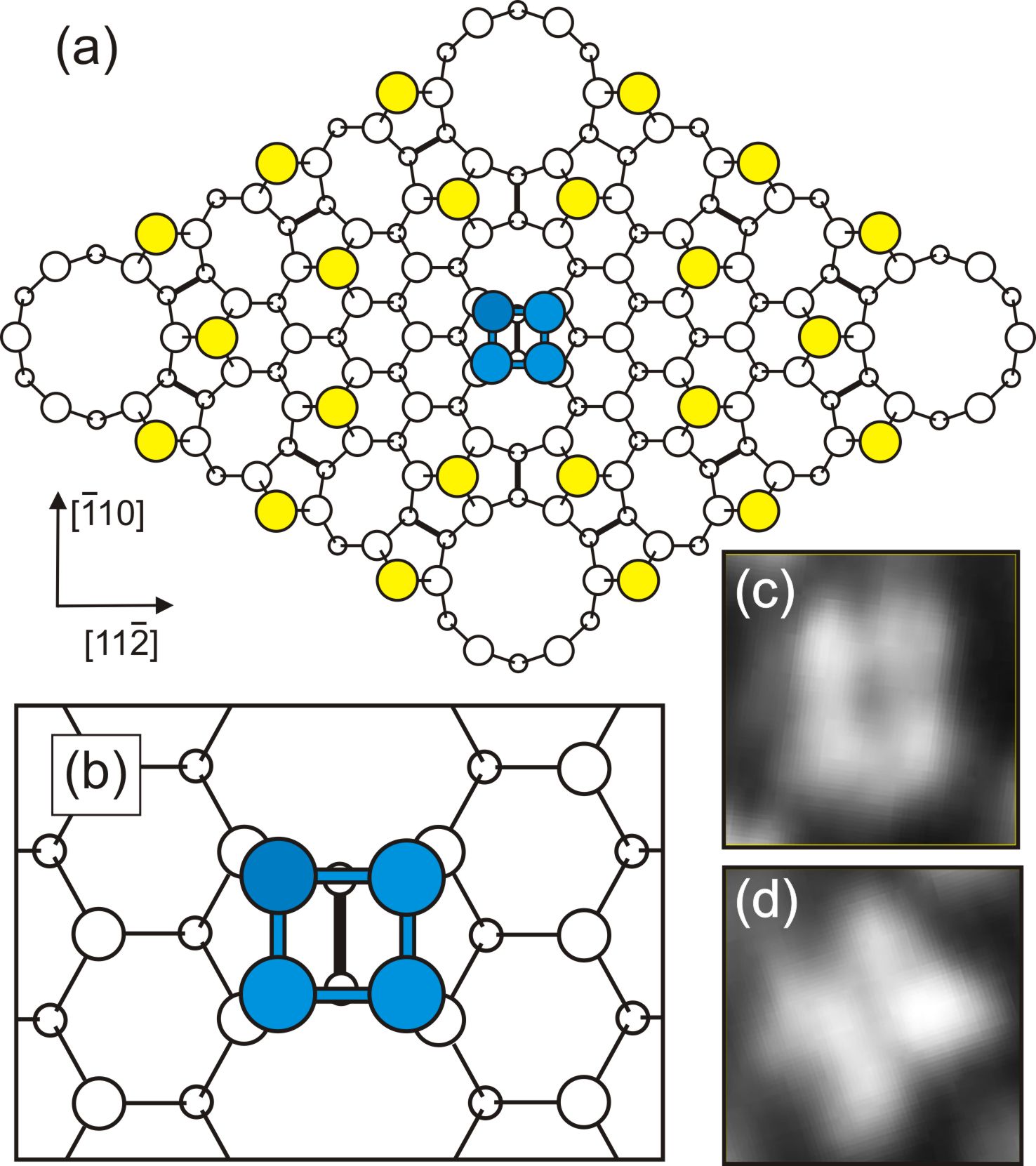}

\caption{\protect\label{fig1}(a) Top view of the Si tetramer model proposed
by Tanaka \emph{et al.} \citep{tan94}. The tetramer formed at the
position of the central dimer of the Si$(111)\textrm{-}7\times7$
cell is shown. Tetramer atoms (blue circles) form a rectangle on the
$7\times7$ structure, yellow circles are adatoms within the DAS structure.
(b) Magnified image of the Si tetramer structure shown in (a). (c),
(d) Experimental STM images ($12\times12$~Å) of the Si tetramer
on the Si$(111)\textrm{-}7\times7$ surface: (c) $U=+1.5$~V, (d)
$U=-2.0$~V. Reproduced with permission from Ref.~\protect\onlinecite{ho04}.
Copyright 2004 Elsevier.}
\end{figure}

We used 6-bilayer thick Si$(111)$ slabs terminated by hydrogen on
the \emph{bottom} side and a $30\,\mathrm{\mathring{A}}$ thick vacuum
layer. The Si atomic positions in the upper face of the slab were
set up according to the atomic structure of the dimer-adatom-stacking
fault (DAS) model of Si$(111)\textrm{-}7\times7$ surface reconstruction
\citep{tak85}, where Si tetramer models, including the one proposed
by Tanaka \emph{et al.} \citep{tan94} (Fig.~\ref{fig1}(a) and (b))
were investigated. The Brillouin zones were sampled using $3\times3\times1$
and $10\times10\times10$ $\mathbf{k}$-point grids for slab (643
atoms) and bulk calculations (2 atoms), respectively \citep{mon76}.
The positions of all slab atoms (except for the Si atoms on the bottom
bilayer and all H atoms) were fully optimized until the atomic forces
became less than $10\,\mathrm{meV/\mathring{A}}$. For the calculation
of the electronic local density of states (LDOS) and STM images, the
basis set employed by the $\mathtt{SIESTA}$ code was optimized as
described in Ref.~\onlinecite{zha18}. Constant-current STM images
were produced within the Tersoff-Hamann approximation using the $\mathtt{WSxM}$
software \citep{hor07}.

\section{Results}

Figures~\ref{fig1}(c) and (d) show experimental STM images of the
Si tetramer formed on the Si$(111)$ surface at the boundary between
two halves of the $7\times7$ unit cell (reproduced from Fig.~1 of
Ref.~\onlinecite{ho04}). The orientation of both images is the same
as in Fig.~\ref{fig1}(a) and (b). The four bright spots at $U>0$
form an approximate squared pattern, whose sides are parallel and
perpendicular to the $7\times7$ boundary where the tetramer resides
(Fig.~\ref{fig1}(c)). However, at $U<0$ the “square” is rotated
by 45$^{\circ}$, thus becoming a “diamond” (Fig.~\ref{fig1}(d)).

The atomic model of the Si tetramer proposed by Tanaka \emph{et al.
}\citep{tan94}, is shown in Figs.~\ref{fig1}(a) and (b). Accordingly,
the tetramer consists of four nearly equivalent Si atoms: two of them
are displaced adatom neighbors from both halves of pristine $7\times7$
DAS unit cell, while two additional atoms should be supplied from
an external source. As it follows from the atomic model, the tetramer
can form either above central or corner dimers of the $7\times7$
DAS structure and this was confirmed experimentally \citep{sat00}.
According to Tanaka \emph{et al.} \citep{tan94}, each tetramer atom
donates some charge to form covalent bonds with neighboring atoms.
Therefore, the high density of empty LDOS is observed at the positions
of Si nuclei (Fig.~\ref{fig1}(c), “square”). The bonds accumulating
the charge exhibit high density of filled LDOS (Fig.~\ref{fig1}(d),
“diamond”). The explanation of experimental STM images of tetramers
proposed by Tanaka \emph{et al.} \citep{tan94} is identical to that
used to justify the formation of symmetric STM images of dimers formed
on Si$(100)\textrm{-}2\times1$ surface at room temperature in early
works (see Ref.~\onlinecite{hat99} for the review of such interpretations).
However, further studies have shown that this picture is invalid.
It was experimentally demonstrated that the STM images of symmetric
dimers are due to their fast flip-flop motion between two stable buckled
states at room temperature, much quicker than the STM time resolution
\citep{hat01}. This conclusion was also supported by numerous theoretical
calculations (see for example Ref.~\onlinecite{dab92}).

\begin{figure}
\includegraphics[clip,width=8cm]{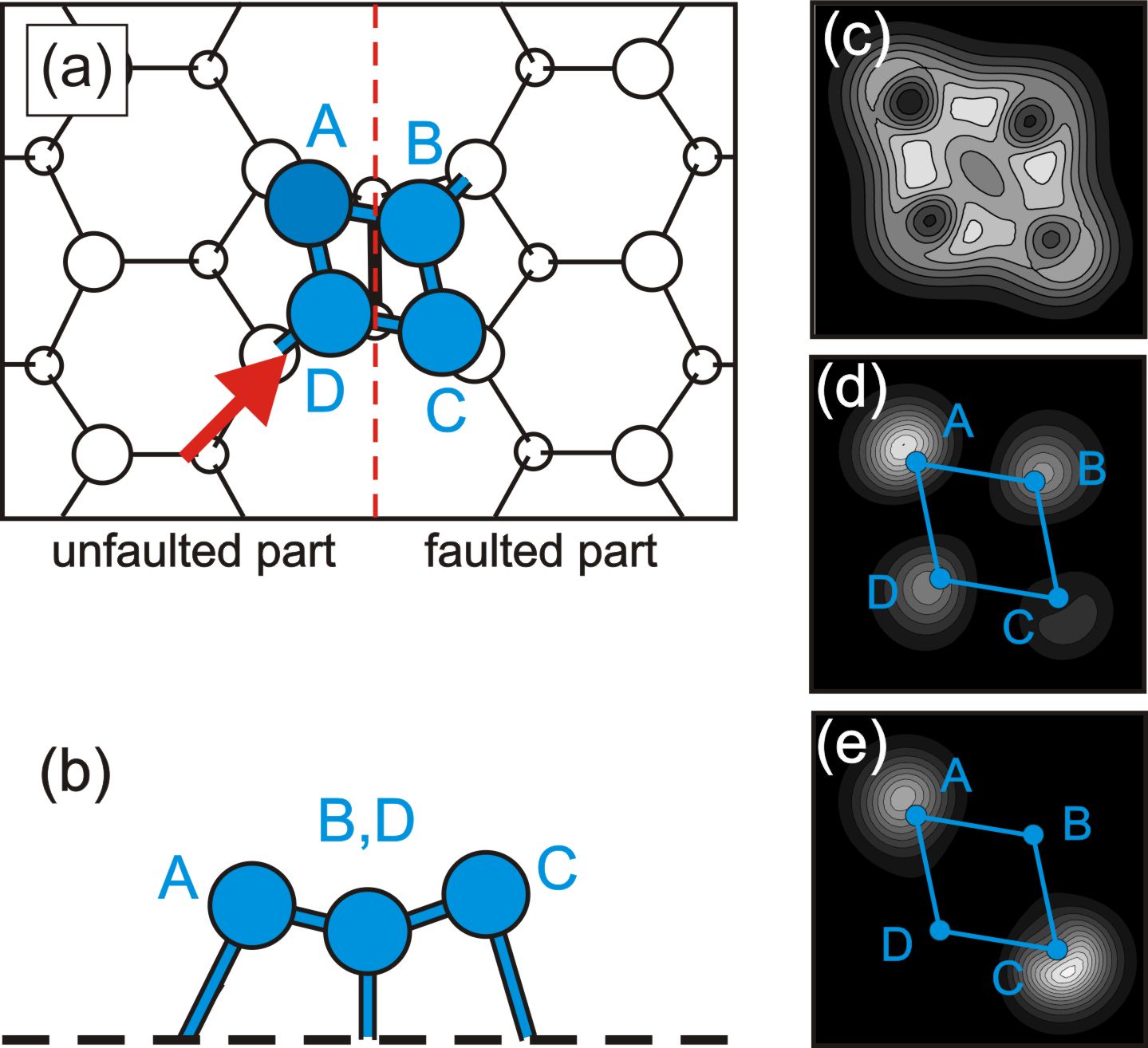}

\caption{\protect\label{fig2}(a) Top view of the Si tetramer buckled atomic
model on the Si$(111)\textrm{-}7\times7$ surface after structure
relaxation. Only one buckled atomic state (among two possibilities)
is shown. The other buckled state is mirror-symmetric to that shown
in (a) with respect to the plane indicated by the red dashed line.
(b) Side view of the Si tetramer in the direction of the red arrow
shown in (a). (c) Horizontal cut of the electron density across the
tetramer atoms. (d), (e) Horizontal cut of the LDOS integrated across
a 1.0~eV window (d) above, and (e) below the calculated Fermi level.}
\end{figure}

Figures~\ref{fig2}(a) and (b) show top and side views of the Si
tetramer atomic model after structure relaxation. The atoms A and
C are pushed away from the geometric center of the tetramer, becoming
more pyramidal ($sp^{3}$-like, the bond angles are: $68.28\text{°}$,
$99.5\text{°}$, $101.6\text{°}$ for A and $66.0\text{°}$, $91.0\text{°}$,
$86.7\text{°}$ for C) ending up with a filled $s$-like dangling
bond. Atoms B and D are pushed inward into a nearly planar configuration
($sp^{2}$-like bonded, the bond angles are: $107.4\text{°}$, $119.9\text{°}$,
$123.6\text{°}$ for B and $106.3\text{°}$, $113.8\text{°}$,
$128.7\text{°}$ for D). Their $p_{z}$-like dangling bonds donate
some electronic charge to lower energy $sp^{3}$-like radical states
in A and C. The net charges on tetramer atoms calculated according
to the Hirshfeld method \citep{gue03} are: $-0.06e$ (A), $-0.07e$
(C), $+0.05e$ (B), $+0.04e$ (D). These figures are close to, although
slightly lower than, the charge transfer of about $0.1e$ from lower
to upper atoms of the buckled dimer on the Si$(100)$ surface as measured
from core-level spectroscopy \citep{rich88}.

The distance between atoms B and D in the relaxed model (2.63~Å)
is close to the calculated bulk Si bond length (2.37~Å). However,
upon inspection of the calculated electron localization function (ELF)
\citep{sil94}, there was no indication of a direct bond between atoms
B and D. For a pair of electrons with little Pauli repulsion effects
such as in a covalent bond, we expect ELF to maximize ($\textrm{ELF}\sim1$),
while in fact it shows a minimum between B and D. This conclusion
is consistent with the electron density between these atoms as depicted
in Fig.~\ref{fig2}(c). Therefore, the main reason for the energy
gain associated with the tetramer buckling is not bond formation between
B and D atoms, but the redistribution of charge from $p_{z}$-like
orbitals of atoms B and D to the $s$-like orbitals of atoms A and
C. The atoms B and D in the relaxed tetramer configuration are about
0.65~Å lower than the highest atom of the tetramer (atom C). The
later (located on the faulted half of the $7\times7$ unit cell) is
also higher than atom A (located on the unfaulted half) by about 0.23~Å
(Fig.~\ref{fig2}(b)).

We calculated the relative energy of tetramers located on top of center
and corner dimers of $7\times7$ reconstruction. We found that the
energy of center tetramer is lower than that of corner tetramer by
0.27~eV. This implies that the center tetramers should form more
frequently than corner tetramers and the formation ratio of the center
to corner tetramer should increase with decreasing the adsorption
temperature. However Sato \emph{et al.} \citep{sat00} observed just
the opposite: at low temperature the corner tetramers were formed
in most cases. This contradiction can be resolved if kinetics of adatom
movement within the $7\times7$ half unit cell is taken into consideration.
According to DFT calculations the most stable adsorption sites for
Si/Si$(111)\textrm{-}7\times7$ and Ge/Ge$(111)\textrm{-}7\times7$
are those close to the corner dimers \citep{cha03,zha23}, and this
is where the adatom was experimentally found to reside most of the
time at low temperature \citep{sat00b}. Therefore, at low temperature
the center tetramer has relatively less chance to form simply because
adatoms avoid these regions. At higher temperature the adsorption
sites closer to the center dimer become populated with traveling adatoms,
thus leading to formation of tetramers on these locations.

Sato \emph{et al.} also observed that when two highly mobile Si atoms
met in neighboring halves of the $7\times7$ cell, they immediately
form the tetramer \citep{sat00}. This observation convincingly shows
that (i) the tetramer consists of four Si atoms: two adatoms from
the $7\times7$ DAS structure and two external adatoms; (ii) the formation
energy of the Si tetramer is lower than that of two separate Si atoms
adsorbed on Si$(111)\textrm{-}7\times7$. We tested the latter premise
by adding separate Si atoms to faulted and unfaulted halves of the
same $7\times7$ unit cell. They were placed in the lowest-energy
adsorption sites, where they form dimers with adatoms of the DAS structure
\citep{cha03,zha23}. We found that the energy gain associated with
the formation of the center tetramer is about 0.29~eV and for the
corner tetramer the gain is only about 0.01~eV.

\begin{figure}
\includegraphics[clip,width=8cm]{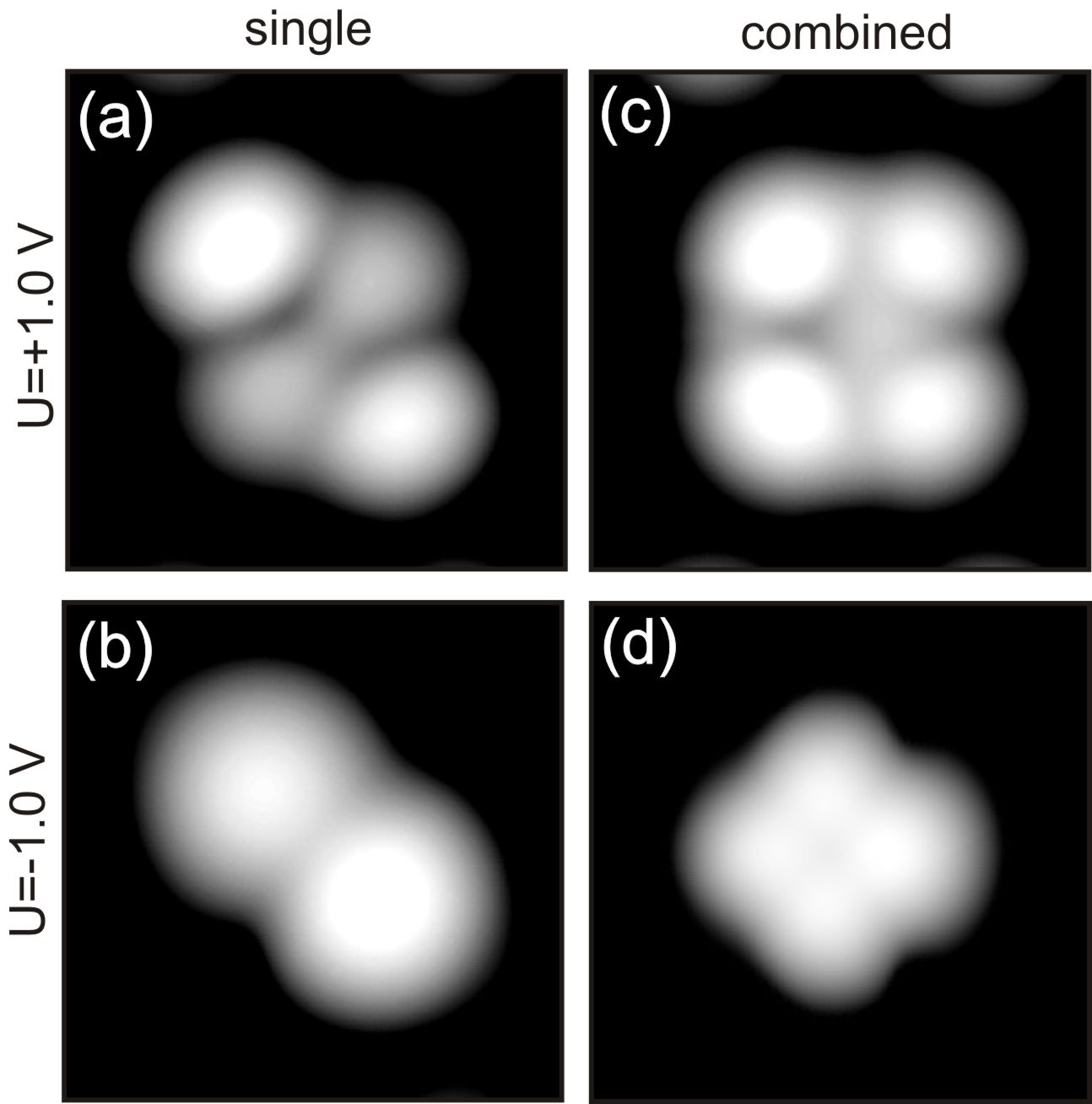}

\caption{\protect\label{fig3}Calculated STM images ($12\times12$~Å) of the
buckled tetramer within the Tersoff-Hamann approximation assuming
a constant current STM operation mode. (a), (c) Empty electronic states,
$U=+1.0$~V; (b), (d) filled electronic states, $U=-1.0$~V. (a),
(b) From the single buckled state; (c), (d) averaged STM images from
the two buckled states. The brightness and contrast of the figures
are not the same and were optimized for best visibility.}
\end{figure}

Figures~\ref{fig2}(d) and (e) show horizontal cuts of LDOS integrated
across a 1.0~eV window (d) above and (e) below the calculated Fermi
level, contributing to the formation of STM images. Empty electronic
states are found on all four atoms of the buckled tetramer. Filled
electronic states are almost exclusively found on atoms A and C, which
is consistent with the atomic charges reported above. According to
the Tersoff-Hamann approximation the STM images are influenced by
both surface topography and the electronic LDOS \citep{ter85}. It
is clear from Fig.~\ref{fig2} that the calculated STM images from
the buckled tetramer will be incompatible with the experimental STM
images shown in Figs.~\ref{fig1}(c) and (d). The calculated STM
images with $U=\pm1.0$~V (presented in Figs.~\ref{fig3}(a) and
(b)) indicate that this is indeed the case.

In order to explain the above puzzle we looked for alternative structures
which could lead to better agreement with the STM data, but also more
stable than those shown in Figs.~\ref{fig1}(a) and (b). A strong
candidate included an additional interstitial atom under the tetramer
structure (not shown here). The resulting atomic geometry is similar
to the pentamer with self-interstitial atom found on $(110)$, $(113)$
and $(331)$ surfaces of Si and Ge \citep{dab94,zha17,zha19,zha20b}.
We found that the interstitial atom indeed stabilizes the tetramer
structure and the buckling was not observed. However, the modified
tetramer model shows three critical flaws: (\emph{i}) formation energy
of the modified tetramer is 0.92~eV higher than that of buckled tetramer
from Fig.~\ref{fig1} {[}calculated from $E_{\textrm{mod}}-E_{\textrm{buck}}-\mu$,
where $\mu$ is the energy per atom in bulk Si, while $E_{\textrm{mod}}$
and $E_{\textrm{buck}}$ are total energies of slabs containing modified
and buckled tetramers, respectively{]}; (\emph{ii}) calculated STM
images of the modified tetramer for positive and negative bias look
alike and similar to the one shown in Fig.~\ref{fig1}(c) (“square”);
(\emph{iii}) the modified model consists of three extra Si atoms,
which is incompatible with experimental observations by Sato \emph{et
al. }indicating two extra Si atoms only \citep{sat00}. Therefore
the Si tetramer model with interstitial atom was rejected and not
considered further.

Another possible explanation for the experimental STM images in Fig.~\ref{fig1}(c)
and (d), would be a thermally-induced flip-flop motion of the tetramer
between two stable buckled states, similar to dimers on the Si$(100)$
surface at room temperature. Figs.~\ref{fig3}(c) and (d) were constructed
from the superposition of calculated STM images obtained from two
symmetry-equivalent buckled states. These images agree well with experimental
counterparts shown in Figs.~\ref{fig1}(c) and (d), indicating that
the dynamic flip-flop motion of Si tetramer may in fact take place.
Indeed, at positive sample bias, the lobes from atoms A and C of the
first tetramer buckled state overlap the lobes from atoms B and D
of the second state (Fig.~\ref{fig3}(a)), and this results in a
“square” shape for the calculated STM image shown in Fig.~\ref{fig3}(c).
For negative sample bias, a nodal plane is observed between the two
large bright spots centered on atoms A and C (Fig.~\ref{fig3}(b)).
When the superposition of STM images from both tetramer buckled states
is constructed (see Fig.~\ref{fig3}(d)), traces of the nodal planes
can still be seen in the form of a faint and dark X-shape pattern.
Four lobes with a diamond shape are formed by the superposition of
LDOS tails from both mirror-symmetric configurations.

The direct experimental observation of tetramer buckling should be
possible at sufficiently low temperature. This would be possible via
detection of random switching of the tunneling current flowing between
tetramer and the STM tip \citep{hat01}. The activation energy barrier
($E_{\textrm{a}}$) for the flipping motion between the two buckled
states of the Si tetramer was calculated using the NEB method as implemented
in the VASP code \citep{kre96,kre96a}. We used five intermediate
images between end-structures, which were relaxed until the maximum
atomic force (subject to the elastic band) was lower than 20~meV/Å.
The ground state configuration as found with VASP (end-structures),
was identical to that obtained with SIESTA. The agreement of both
codes regarding the description of several Si surfaces, has been demonstrated
elsewhere \citep{zha22}.

The linear distance $\Delta R$ separating initial ($r_{i=0,k}$)
and final ($r_{i=6,k}$) atomic geometries, here identified as images
$i=0$ and $i=6$, is obtained from $(\Delta R)^{2}=\sum_{k}(r_{i=6,k}-r_{i=0,k})^{2}$
with $k=1,\ldots,3N$ and $N$ being the total number of atoms. For
the present problem we have $\Delta R=1.7$~Å only (corresponding
to an average 0.3\,Å step size), thus justifying the use of 5 intermediate
images.

\begin{figure}
\includegraphics[clip,width=8cm]{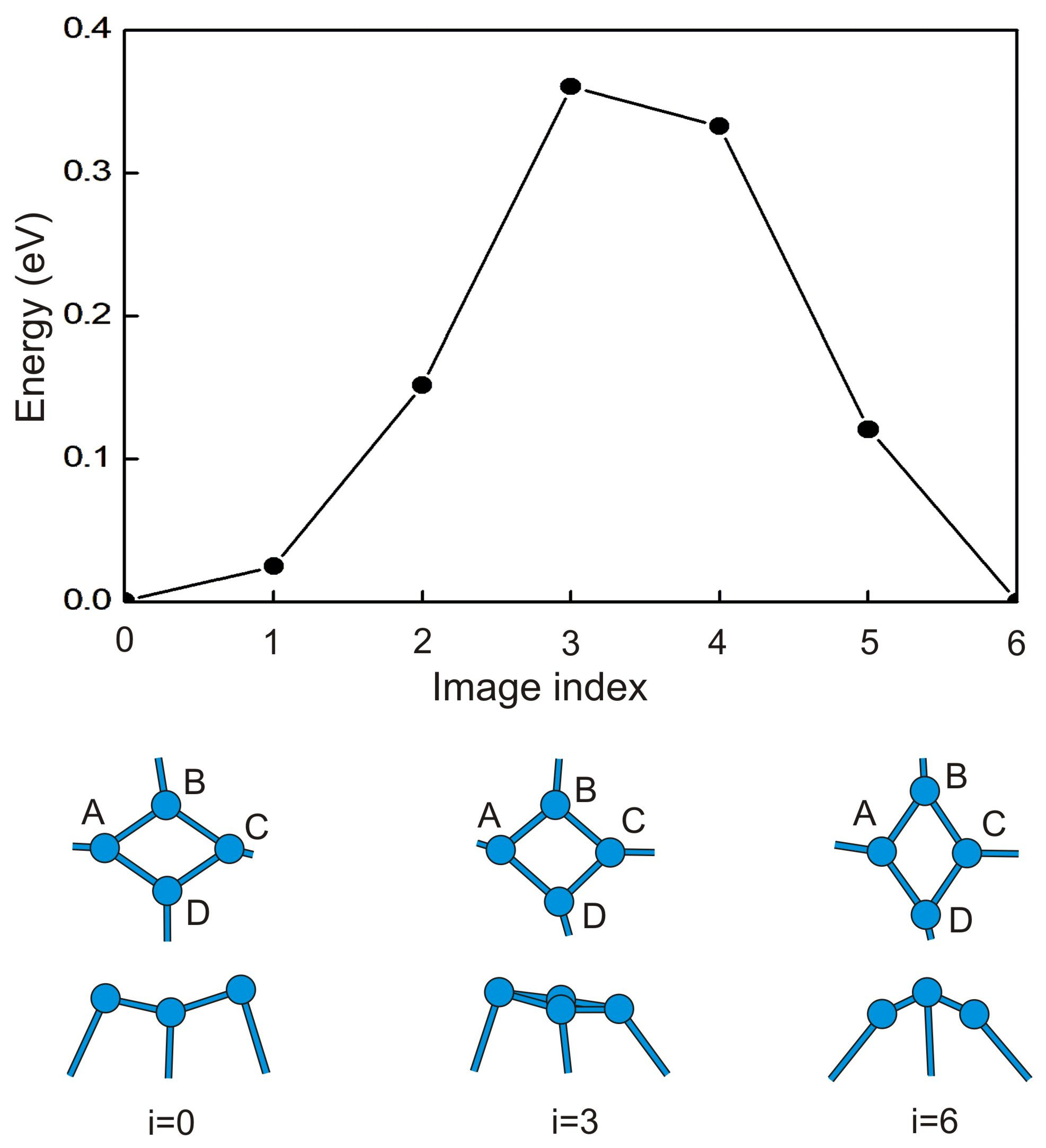}

\caption{\protect\label{fig4}Calculated minimum energy path along the transformation
between two equivalent energy minima of the Si tetramer on Si(111)
surface. The calculations employed the NEB method. Dots in the graph
represent the energies of intermediate supercell images ($i=1,\ldots,5$)
with respect to those of initial ($i=0$) and final ($i=6$) configurations.
At the bottom, we represent top views (upper insets) and side views
along the red arrow of Fig.~\ref{fig2}(a) (lower insets), of local
geometries of initial image (left), saddle-point (middle), and final
image (right).}
\end{figure}

The results from the NEB calculations are shown in Fig.~\ref{fig4},
where we also depict the local geometries of both end-images and saddle
point. The results display an asymmetric energy profile, which may
be explained in the following way: The initial guess for the mechanism
(a sequence of linearly interpolated images between end-structures)
corresponds to a straight path that goes over a local maximum in the
potential energy landscape. During the NEB relaxation the elastic
band slips downhill until it reaches a nearby saddle point, essentially
circumventing the initial local maximum. We understand the result
as a two-step mechanism, starting with a climbing stage involving
a transfer of the $sp^{3}$ hybrid state localized on atom C (see
left inset of Fig.~\ref{fig4}) into atom A in the opposite apex
of the tetramer structure at the saddle-point (middle inset). Note
that atoms A and C sit on top of unfaulted and faulted halfs of the
Si(111)-$7\times7$, and so they are not equivalent. The second half
of the mechanism is a downhill process, involving another \emph{jump}
of the $sp^{3}$ hybrid state on atom A in the saddle-point, to its
adjacent atom B (see right inset of Fig.~\ref{fig4}). This is accompanied
by the realignment of the tetramer configuration to reach the final
image.

We found that the activation energy for the minimum energy path is
$E_{\textrm{a}}\thickapprox0.36$~eV, which typically has an error
bar of $\sim\!0.1$~eV, There are studies which point to transition
barriers involving $sp^{3}\rightarrow sp^{2}+p$ hybridization changes
on Si surfaces as low as few meV \citep{Bra18}. However, there are
also works indicating larger figures. For instance, the barrier for
the flipping of dimers on Si(100) has been calculated as about 0.2~eV
\citep{Paz01,Hea01}, in line with the observed phase transition between
symmetric and asymmetric dimers at 120~$^{\circ}\textrm{C}$, allowing
to consider our results as reasonable. We can only conclude that the
barriers must depend on the local environment of the flipping Si structures.

It is however important to note that despite its recognized reliability,
the NEB method scans a limited fraction of the phase-space, and therefore
we cannot guarantee that the result corresponds to the true saddle-point.
In that regard, we must take the calculated height of the barrier
as an upper bound of the real barrier.

It is possible to estimate the sample temperature threshold ($T_{\textrm{c}}$),
below which the tetramer should be frozen in one buckled state (within
the timescale of a STM measurement). To this end, we make use of the
frequency ($f$) of thermally activated switching of the buckled state
using an Arrhenius relation $f=f_{0}\exp\left(-E_{\textrm{a}}/k_{\textrm{B}}T_{\textrm{c}}\right)$,
where $k_{\textrm{B}}$ and $f_{0}$ stand for the Boltzmann constant
and attempt frequency, respectively, the later being approximated
by the Debye frequency of Si (14~THz). Assuming ``frozen'' tetramers
as those flipping slower than 1~Hz, we find $T_{\textrm{c}}\approx140$~K.
According to the experimental results by Sato \emph{et al.} \citep{sat00}
and Ho\emph{ et al.} \citep{ho04} the STM images of tetramers look
very similar at room and at liquid nitrogen temperature (77~K). This
indicates that the Si tetramers should perform dynamic buckling even
at the temperature below $T_{\textrm{c}}$.

Among possible reasons for a high-frequency flip-flop motion at $T=77$~K
we could have the softening of the barrier height due to charge injection,
or due to the electric field from the STM tip. The first possibility
was investigated by adding/removing one electron to/from the supercell
hypothesizing that the tetramer could create an electron/hole trap
at the surface, and by that, to change the local bonding. However,
from NEB calculations we found no charge effects whatsoever.

The impact of the electric field is also expected to be no more than
few meV. This was estimated from a dipole induced charge of $\approx\!0.1e$
on tetramer atoms, $\approx\!0.5$~Å height difference between them,
and an electric field created by the STM tip at a distance of $\approx5\,\mathrm{\mathring{A}}$
from the surface with $U\approx1\,V$.

It is also unlikely that vibrational entropy change across the transformation
barrier (incorporated in the vibrational free energy change) could
contribute with more than about $\Delta F\approx50$~meV to the barrier
at $T=77$~K. This figure arises from an estimate, where the difference
between vibrational free energies of ground state and saddle-point
structures, boils down to

\begin{equation}
\Delta F=k_{\textrm{B}}T\ln\left[2\sinh\left(\frac{\hbar\omega_{\mathrm{D}}}{2k_{\textrm{B}}T}\right)\right],
\end{equation}
which accounts for a single effective vibration whose normal mode
is directed along the transformation path, the respective frequency
is of the order of the Si Debye frequency ($\hbar\omega_{\textrm{D}}=58$~meV),
and $\hbar$ is the reduced Plank constant. The magnitude of $\Delta F$
is in line with previous and more accurate calculations of such effects
\citep{Bra20}. Of course, this simple model assumes a minute difference
between the the two geometries and respective sets of vibrations involved.
Hence, any definite statement requires the explicit evaluation of
the free energy change and of the dynamical matrix for the atoms in
the slab \citep{Bra20}.

An alternative explanation, which unfortunately we cannot provide
a quantitative assessment, is based on scattering of tunneling electrons.
In this case, tunneling electrons involved in the measurement could
excite phonons localized on the tetramer (local heating), thus helping
the diamond-like structure to overcome the energy barrier between
two buckled states. This effect has been invoked to interpret variable-temperature
scanning tunneling microscopy experiments on Si(001), where the amount
of $P(2\times1)$ surface reconstruction area measured at $T=65$~K,
strongly depended on the tunneling current \citep{Mit00}. Current-induced
scattering was also suggested as the most probable reason to justify
the observation of symmetric Si dimers on Si$(100)\textrm{-}2\times1$
at low temperature \citep{kaw04}.

\section{Conclusions}

In conclusion, the atomic structure of Si tetramers formed at Si~/~Si$(111)\textrm{-}7\times7$
homoepitaxy, was investigated by means of first-principles calculations
based on density functional theory. It is shown that the atomic model
of Si tetramer, consisting of a four-member ring of Si atoms, describes
well the experimental STM images, only if the tetramer is assumed
to dynamically buckle between two equivalent states with equal contribution
to the resulting STM images. This effect nicely explains the change
between squared and diamond shaped STM features upon switching between
positive and negative biased measuring conditions.
\begin{acknowledgments}
R.Z. and J.C. thank the FCT through projects Refs: LA/P/0037/2020,
UIDB/50025/2020, UIDP/50025/2020. D.S. thanks the Russian Science
Foundation (Project No. 19-72-30023) for financial support. The authors
thank the Novosibirsk State University for providing computational
resources, as well as the RNCA for providing computational time on
the Oblivion supercomputer through project 2024.04745.CPCA.A1.
\end{acknowledgments}

\section*{Data availability}

The data that support the findings of this study are available from
the corresponding author upon reasonable request.

\bibliographystyle{apsrev4-1}
\bibliography{refs}

\end{document}